\documentclass[reprint,aps,prl,showpacs]{revtex4-1}
\usepackage{graphics,graphicx,amssymb,amsmath,bm,dcolumn,url}
\begin{document}
\title{A mechanism for stickness, dealing with extreme events}
\author{Taline Suellen Kruger$^1$}
\author{Paulo Paneque Galuzio$^1$}
\author{Thiago de Lima Prado$^1$}
\author{Sergio Roberto Lopes$^{1}$}
\email{lopes@fisica.ufpr.br} 
\author{Jos\'e Danilo Szezech Jr$^2$}
\author{Ricardo Luiz Viana$^1$}
\affiliation{$^1$Departamento de F\'{\i}sica, Universidade Federal do Paran\'a, Curitiba, PR, Brazil\\
$^2$ Departamento de Matem\'atica e Estat\'{\i}stica, Univ. Est. de Ponta Grossa, Ponta Grossa, Paran\'a, Brazil}

\date{\today}  

\begin{abstract}

In this letter we study how hyperbolic and non hyperbolic regions in the neighborhood of a resonant island perform a important role allowing or forbidding stickiness phenomenon around islands in conservative systems. The vicinity of the island is composed by non hyperbolic areas that almost prevent the trajectory to visit the island edge. For some specific parameters there are tiny channels embedded in the non hyperbolic area that are  associated to hyperbolic fixed points present in the neighborhood of the islands. Such channels allow the trajectory to be injected in the inner portion of the vicinity. When the trajectory crosses the barrier imposed by the non hyperbolic regions, it spends a long time to abandon the surrounding of the island, since the barrier also prevents the trajectory to scape from the neighborhood of the island. In this scenario the non hyperbolic structures are the responsible for the stickiness phenomena, and more than that, the strength of the sticky effect. We reveal that those properties of the phase space allow us to manipulate the existence of extreme events (and the transport associated to it) responsible for the non equilibrium fluctuation of the system. In fact we demonstrate that monitoring very small portions of the phase space (namely $\approx 4\times 10^{-4}$ \% of it) it is possible to generate a completely diffusive system eliminating long time recurrences that result from the stickiness phenomenon.
\end{abstract}

\date{\today}
\maketitle

Understanding the transport properties of Hamiltonian system is one of the major objective in the statistical analysis of dynamical systems \cite{zaslavsky_livro,lichtenberg}. Recent works have shown that for a large class of systems, including those presenting mixed phase space, the transport can not be treated considering just ergodic theory or random phase approximation \cite{zaslavsky_2002}. One class of such systems is the low dimensional (3/2 or 2 degrees of freedom) Hamiltonian system (1/2 degree of freedom corresponds to a periodical disturbance). 
Low dimensional Hamiltonian systems, commonly present nonuniform phase spaces composed by regular (islands) and chaotic regions. The interface between these regions is far from being a smooth surface and the dynamics near the edge between chaotic and regular regions is very complex and have been not well understood so far. The complexity comes mainly due to the presence of stickiness in the boundaries of islands \cite{meiss_1985}. The sticky effect forces a trajectory injected into the boundary area to stay near the boundary for long periods of time. One of the main  consequences  of this phenomena is the existence of power law tails in the Poincar\'e recurrence times making the system to present distribution of recurrence times displaying algebraically decay for long times rather than a exponential decay as expected for a normal transport system \cite{venegeroles_2009,zaslavsky_2002,altmann_2008,sanjuan_2008,
altmann_2010,tamas_lai_book, alus_2014}. In such a situation these systems are characterized as out of equilibrium. 

Some features of the kinetics of Hamiltonian systems are important to understand anomalous transport and super diffusion. The phase space topology of theses systems plays a crucial role in the anomalous transport and in sticky phenomena \cite{meiss_1993,lopes_2012}. Many problems of science such as particle advection in fluids \cite{balbiano_1994,tel_2005}, transport in plasma fusion devices \cite{lichtenberg,szezech_2012}, celestial mechanics \cite{Efthymiopoulos_1999} and many others found applications in stickiness occurrences. 
 
Previous works have studied properties of the boundaries between regular and chaotic regions focusing mainly in the role of the stickiness in the dynamics \cite{zaslavsky_2002a,bunimovich_2012} and also the cantori structures derived from the break of tori \cite{mackay_1984}. When the system presents stickiness, the dynamics of orbits in chaotic sea is observed to be intermittent, where after periods of chaotic motion away from the influence of a sticky island, the system presents periods of almost regular motion.  However a mechanism based on the topology of the island vicinity and how that topology affects an injected and subsequently ejected  trajectory from the sticky area is not completely known. 

Here we study how characteristics of the topology of the system, namely hyperbolic and non hyperbolic regions on the neighborhood of a resonant island perform an important role in order {\it to establish the presence and more than that, the strength of the sticky effect}. We show that the sticky effect is associated with the presence of injection channels related to the crossing of stable and unstable manifolds of hyperbolic fixed points in the vicinity of the island, allowing the trajectories to shift between sticky and nonsticky areas of the phase space. We show that the effectiveness of such channels to capture trajectories to the sticky area is closely related to the degree of hyperbolicity of the close surrounding area of hyperbolic fixed points located in the neighborhood of an island.  Finally we make use of the presence of the hyperbolic channels to avoid (control) extreme recurrence events and the anomalous transport associated to it, resulting from a trajectory injection into the sticky area.

Here we characterize a hyperbolic region of the phase space ${\cal S}$ as an ensemble for which the tangent phase space splits continuously into stable (SM) and an unstable (UM) manifolds. SM and UM  are invariant under the system dynamics: infinitesimal displacements in the stable (unstable) direction suffer exponentially decay as time goes forward (backward) \cite{guckenheimer_2002}. In addition, it is required that the angles between the stable and unstable directions to be uniformly bounded away from zero. In this way, in order to quantify the degree of non hyperbolicity related to the phenomena we describe in this letter, let us consider an initial condition $(p_0,x_0)$ and an unit vector ${\boldsymbol v}$, whose temporal evolution is given by  
\begin{equation}
{\boldsymbol v}_{n+1}={\cal J}(p_n,x_n){\boldsymbol v}_n/||{\cal J}(p_n,x_n){\boldsymbol v}_n||,
\end{equation}
 where ${\cal J}(p_n,x_n)$ is the Jacobian matrix of the  map. For $n$ large enough, ${\boldsymbol v}$ is parallel to the Lyapunov vector ${\boldsymbol u}(p,x)$ associated to the maximum Lyapunov exponent $\lambda_u$ of the map orbit starting by $(p_0, x_0)$. A backward iteration of the same orbit gives us a new vector ${\boldsymbol v}_n$ that is parallel to the direction ${\boldsymbol s}(p,x)$, the Lyapunov vector associated to the minimum Lyapunov exponent $\lambda_s$ \cite{grebogi_1990}. 
For regions where $\lambda_s < 0 < \lambda_u$ the vectors ${\boldsymbol u}(p,x)$ and ${\boldsymbol s}(p,x)$ are tangent to the UM and SM, respectively, of a point $(p,x)$. 

The (non)hyperbolic degree of a region ${\cal S}$ can be studied computing the local angles between the two manifolds 
\begin{equation}
\theta(p, x) = \cos^{-1} (| {\boldsymbol u}\cdot {\boldsymbol s}|),
\label{theta}
\end{equation}
for $(p, x) \in {\cal S}$ \cite{ginelli_2007}. So, $\theta(p,x) \sim 0$ denotes tangency between UM and SM  at $(p,x)$.  The general method used to calculate the $\theta$ angles follows the reference \cite{ginelli_2007}.   

Chaotic orbits of two dimensional mappings are often non hyperbolic since the SM and UM are tangent in infinitely many points. As an illustration of this effect we consider a periodically kicked rotor subjected to a harmonic potential function - the Chirikov-Taylor map \cite{lichtenberg}, whose dynamics is two dimensional. The dynamics of a periodically kicked rotor can be described in a periodic phase space $[-\pi, \pi)\times [-\pi,\pi)$, whose discrete-time variables $p_n$ and $x_n$ are respectively the momentum and the angular position of the rotor just after the $n$th kick, with the dynamics given by the following equations:  
\begin{eqnarray}
p_{n+1} &=& p_n + K \sin(x_n), {\rm mod}\, 2\pi \\
x_{n+1} &=&  x_n + p_{n+1}, {\rm mod}\, 2\pi
\label{map}
\end{eqnarray}
where $K$ is related to the kick strength.

\begin{figure}
\includegraphics[width=\columnwidth]{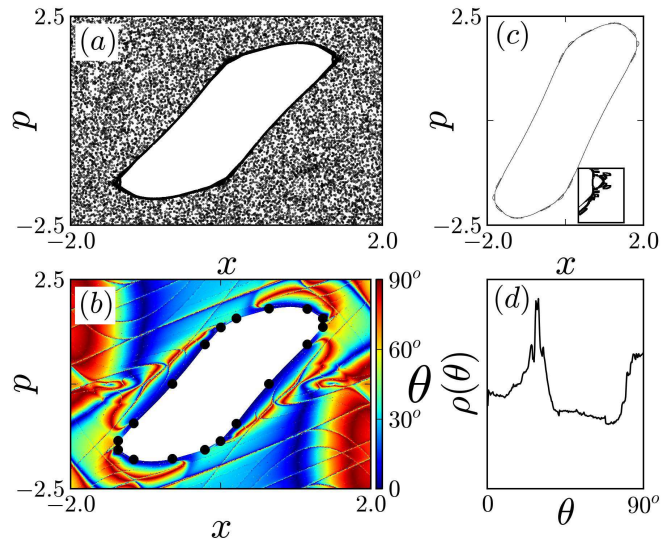}
\caption{\label{sticki} (color online) (a) Phase space for the kicked rotor map, darker regions around the main island reflects the effect of the stickiness. (b) Phase space distribution of angle between unstable and stable manifold, dark (blue) tones mean strong non hyperbolicity. (c) An example of the edge detection algorithm used here to compute the vicinity of the island. In the inset we display a magnification showing details of the edge.  (d) Probability distribution function of the angle between stable and unstable manifolds, $\rho(\theta)$.}
\end{figure}

In order to exemplify the dynamics of the Chirikov-Taylor map and its sticky phenomena,  Fig. \ref{sticki}-a presents a portion of the phase space for a typical trajectory of the system for $K=3.0$. The denser areas near the island result from the sticky effect due to the time the trajectory remains near the edge of the island. In order to characterize the hyperbolicity of the surrounding areas of the island Fig. \ref{sticki}-b displays the local angle between stable and unstable manifolds ($\theta$), Eq. (\ref{theta}), near the resonant island. It is clear that the major part of the vicinities of the island is composed by strong nonhyperbolic areas, represented by dark blue areas in Fig. \ref{sticki}-b (tangencies between UM and SM). A small fraction of the vicinity is observed in  red -- yellow -- green tones and corresponds to angles greater than $30^0$ and responsible for the weak hyperbolic part of the vicinities. The role of both areas will be clear later on in the text. In Fig \ref{sticki}-b we also define the angle $\phi = \arctan(p/x)$, defined in the interval $[-\pi/2, +\pi/2]$, so each point in the vicinity of the island can be identified by a single number. The $9+9\, (\rm symmetric)$ hyperbolic fixed points present in the neighborhood of the main island are identified as black bullets in Fig. \ref{sticki}-b. In order to make clear what we call the vicinity of the island, Fig \ref{sticki}-c displays the result of our algorithm for edge island detection, details magnified in the inset. 
Finally in Fig \ref{sticki}-d we plot the probability distribution function of the angles between SM and UM $\rho(\theta)$, so $\rho(\theta)d\theta$ represents the probability  to find a angle between $\theta$ and $\theta + d\theta$ in the phase space ensemble displayed in Fig. \ref{sticki}-b. The large plateau for small $\theta$ angles reflects the strong nonhyperbolic character of the region.

\begin{figure}
\includegraphics[width=\columnwidth]{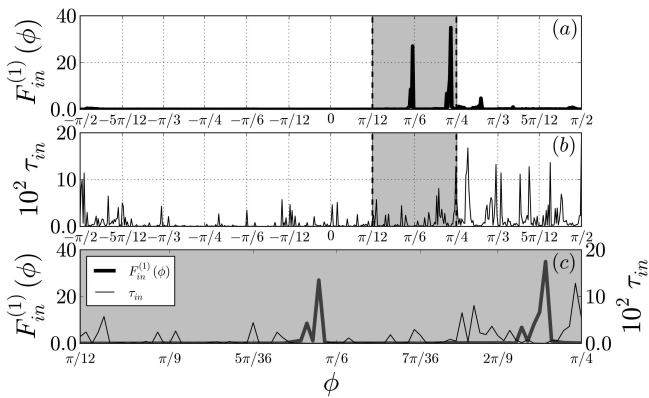}
\caption{\label{distrib} (a) Probability distribution function of income angle in the vicinity of a island of the Chirikov-Taylor map. (b) Average time spent in the sticky area as a function of the injected angle.(c) Magnification of panel (a) and (b) near the maxima of $F_{in}^{(1)}(\phi)$}
\end{figure}

The topological properties of the phase space in the vicinities of an island play an important role in the sticky mechanism. In order to explore the relation between topological characteristics  of the phase space and the way trajectories visit an island vicinity and stick to it, we define the probability density function $F_{in}^{(1)}(\phi)$ for trajectories injected into the vicinities of  the island considering the vicinity computed by our algorithm (Fig. \ref{sticki}-c). $F_{in}^{(1)}(\phi) d\phi$ is the probability that a typical chaotic trajectory to visit the vicinity of the island through a angle between $\phi$ and $\phi+d\phi$. In Fig. \ref{distrib} we plot $F_{in}^{(1)}(\phi)$, panel (a) as well as the average time the trajectory remains in the vicinity of the island when injected by a particular angle, panel (b). In panel (c) we magnify the gray region of panel (a) and (b). 
Observe that although the major part of the trajectories are injected by just few angles inside the red-yellow-green tones regions around the island in Fig. \ref{sticki}-a  (weak hiperbolic regions ), these specific trajectories spend, in average, a short time mapping the sticky area. Trajectories are easily injected into the sticky areas by weak hyperbolic areas surrounding the island but almost all of them are also easily ejected from it. Those trajectories do not contribute for the phenomena of stickiness and do not make any substantial changes in the Poincar\'e recurrence time for the dynamics.   

\begin{figure}
\includegraphics[width=\columnwidth]{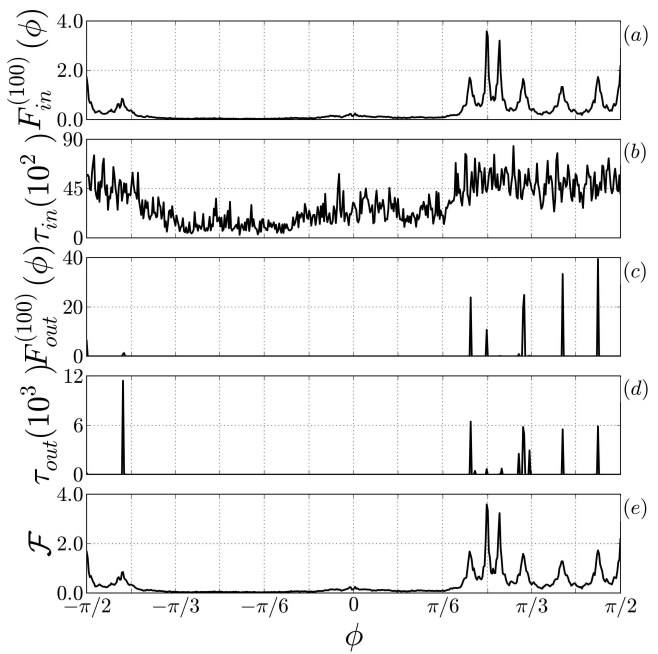}
\caption{\label{distrib100} (a) Probability distribution function of income angle into the vicinity of the main island of the Chirikov map. (b) Average spent time in the sticky area as a function of the injected angle. (c) Probability distribution function of outcome angle from the vicinity of the main island of the Chirikov map. (d) Average spent time in the sticky area as a function of the ejected angle. (e) Probability distribution of injected angle, subjected to the condition that the time of stickiness to be greater than 1000 iterations}
\end{figure}

In order to distinguish sticky trajectories from those that just reach the island edge and leave it quickly, we compute the probability density function $F_{in}^{(100)}(\phi)$, of trajectories injected into the sticky area by a specific angle {\it considering that once a trajectory reaches the vicinity of the island, it remains mapping the same set of points as computed by our algorithm of island edges detection for at least $100$ iterations}. We identify such  trajectories as sticky ones.  The result is plotted in Fig. \ref{distrib100}-a. Almost all sticky trajectories are injected in very specific intervals of angles. Each angle intervals are directly related to the angular location of the chain of periodic points (set as black bullets in Fig. \ref{sticki}-c). It is possible to conclude that the trajectories are injected into the sticky area when they are tangent to the stable manifold of the $18^{th}$ order fixed points chain located in the vicinity of the main island.
All trajectories that are not tangent enough to the stable manifold of the hyperbolic fixed point do not cross the tiny hyperbolic channel produced by the crossing between stable and unstable manifolds of the fixed point and can not be injected into the stick area.

Considering those trajectories injected into the stickiness remaining mapping the edge for at least $100$ iterations, we compute in Fig. \ref{distrib100}-b the average time they spend near the island (sticky trajectories) as a function of the injected angle. 
In Fig.  \ref{distrib100}-c we graph the probability distribution function of sticky trajectories as a function of the ejected angle $F_{out}^{100}(\phi)$. It is clear the almost discrete nature of the distribution. All ejected trajectories follow the unstable manifold of the hyperbolic fixed points moving along a narrow channel departing from the fixed point. We plot in Fig. \ref{distrib100}-d the average time the trajectories stay in the sticky region (at least $100$ iterations inside the sticky area) as a function of the outcome angle. From Fig. \ref{distrib100}-d we can conclude that sticky trajectories are ejected only by few angle intervals $\phi_{out}=\phi({\rm max}(\tau_{out}))$. Therefore, we are able to calculate the probability $\mathcal{F}(\phi)d\phi$ that a given trajectory will enter the sticky region considering only trajectories that leave these region through the angle $\phi_{out}$. The result is plotted in Fig. \ref{distrib100}-e. The great similarity between $\mathcal{F}$ and $F_{in}^{(100)}$ suggests that the or previous conclusions are consistent. Therefore, we can argue that the local maximum of $F_{in}^{(100)}$ represent the sticky angles, {\it i. e.}, the angles that once a trajectory is injected from one of them, there is a great probability that this trajectory turn to be stuck to the island. These maxima correspond to the same region where are located hyperbolic points, confirming the hypotheses that these points provide channel for a typical trajectory to enter in the sticky region. 
Figs. \ref{distrib100}-a and \ref{distrib100}-e, clearly show us that both figures are almost identical, a strong suggestion that all trajectories leave the sticky regions by the hyperbolic channels departing from the hyperbolic fixed points. 

\begin{figure}
\includegraphics[clip,width=\columnwidth]{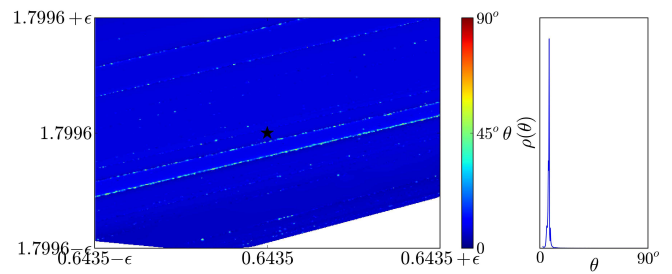}
\caption{\label{f_x_theta} (color online) (a) Degree of hyperbolicity of the vicinity ($\epsilon=0.0025$) of one of the eighteen fixed points present around the main island of the map \ref{map}. (b) $\rho(\theta)$ characterized by just on maximum when  computed near a fixed point around the main island}
\end{figure}

\begin{figure}
\includegraphics[clip,width=\columnwidth]{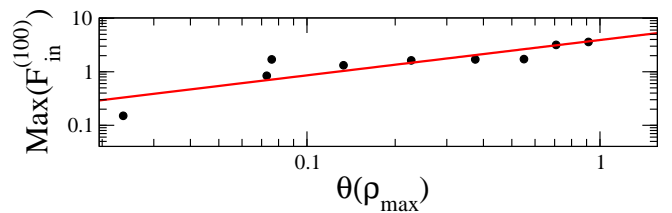}
\caption{\label{rho_imperbolic} (color online) The effectiveness of the injection channels measured using the $\max F^{100}_ {in}$ as a function of the angle between UM and SM.}
\end{figure}

The presence of stickiness around an island is predicted by some theory \cite{meiss_1985, bunimovich_2012}  but an analysis of Fig. \ref{distrib100}-a shows that the effectiveness of a trajectory injection or ejection by a particular hyperbolic channel is not the same for all fixed points, as can be observed by the different amplitude of maxima of $F_{in}^{(100)}$ in Fig. \ref{distrib100}-a. In order to make clear the role of the hyperbolicity of the close vicinity of the hyperbolic fixed points in the injection and ejection phenomena of sticky trajectories, firstly, we present as an example,  in Fig. \ref{f_x_theta}-a the degree of hyperbolicity of one of the $18$ fixed points presents around the main island of the Chirilov-Taylor map. As observed in Fig. \ref{f_x_theta}-b the probability density function $\rho(\theta)$ presents just a sharp maximum due to the almost unique angle between stable and unstable manifold computed in the close vicinity of the fixed point. All other fixed points present similar sharp peaks in the probability density function $\rho(\theta)$, nevertheless each hyperbolic point has its own angle for the maximum of $\rho(\theta)$ characterizing its own {\it degree of hyperbolicity}.  Secondly, to demonstrate the relation between the degree of hyperbolicity of the close vicinity of the fixed points, and the effectiveness of the hyperbolic channels related to each fixed point to capture sticky trajectories,  we graph in Fig. \ref{rho_imperbolic} the maxima of the function $F_{in}^{100}(\phi)$ as a function of the  degree of hyperbolicity measured by the angle for which the function $\rho(\theta)$ presents a maximum $\theta(\rho_{\rm max})$. The red line is a power law fitting that serves  us as eye guide. The result presented in Fig. \ref{rho_imperbolic} clearly shows that the effectiveness of the channels is a function of the degree of hyperbolicity of the close region of the fixed points. Small values of  $\theta(\rho_{\rm max})$ is related to the fact that just a very small portion of the surrounding area of the fixed point is occupied by the injection/ejection hyperbolic channel. As a result the function $F_{in}^{(100)}(\phi)$ presents a relatively small maximum, meaning that just a small fraction of trajectories can cross the channel in an injection or ejection process from the sticky area. 

\begin{figure}
\includegraphics[clip,width=1\columnwidth]{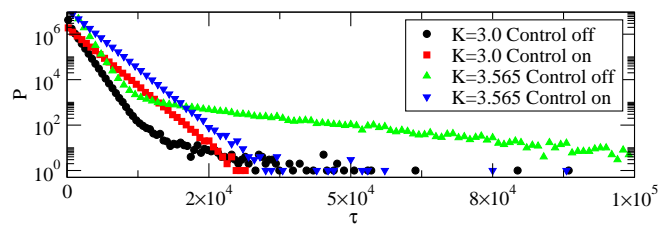}
\caption{\label{rec_poicare} (color online) Recurrence time for the Chirikov-Taylor map with and without control}
\end{figure}

Such properties of the phase space allow us to manipulate the non equilibrium fluctuation of the system. To show that it is possible to control the non equilibrium fluctuations that arise due to the presence of stickiness, we track the position of the trajectory and once it maps a small circle of radius $0.003$ centered in one of the 18 fixed points we perturb the trajectory, so a possible crossing of the channel and consequent stick of the trajectory is avoided. Numerically, we perform a restart of the trajectory outside the injection channel. Results for the Poincar\'e recurrence time for the system with and without the control mechanism for two values of $K$, $K=3.0$ and $K=3.565$ (a large stickiness case) are plotted in Fig. \ref{rec_poicare}. Black bullets and green up-triangle display the time distributions for Poincar\'e recurrence without any control mechanism. As can be observed for large recurrence time, a strong fluctuation of the exponential law is observed. In fact for large recurrence time the distribution has a power law decay as a result of the sticky phenomena in the recurrence time. The time distribution of the Poincar\'e recurrence time for the  system subjected to our control mechanism is displayed red squares and blue down-triangle. For this case, almost all fluctuation for long recurrence time is absent, corroborating to the idea that all nonequilibrium fluctuation in the system is now absent since the stickiness is avoided. Additionally observe that the exponential rate for both $K$ values is the same supporting the idea that the behavior of the system is now completely diffusive independently of the $K$ value.  All results here are presented for two values of $K$ but similar results are obtained for other values of the nonlinear parameter.

In conclusion, we describe in this letter a mechanism to suppress the effect of stickiness based on the knowledge of the nonhyperbolic structure on the edge of a island of a Hamiltonian 2 degree of freedom system. We show that the  effectiveness of an island edge to stick trajectories is directly related to the degree of hyperbolicity of small areas surrounding fixed points around the island. We show that monitoring those areas of the phase space, it is possible to generate a complete diffusive processes without (almost) any influence of the large recurrence time due to the stickiness phenomena. Since very large sticky times make the system to present extreme events in the dynamics, it possible to affirm that, once under control, we can turn the out-of equilibrium system into a in equilibrium one.

This work has partial financial support from CNPq (PROCAD), CAPES, and Funda\c c\~ao Arauc\'aria (Brazilian agencies). Computer simulations were performed at the LCPAD cluster at Universidade Federal do Paran\'a, supported by FINEP (CT-INFRA).


\begin{thebibliography}{99}
\bibitem{zaslavsky_livro} G. M.  Zaslavsky {\it Hamiltonian Chaos and Fractional Dynamics} (Oxford University Press Inc, New York. 2005). 
\bibitem{lichtenberg} A. J. Lichtenberg, and M. A. Lieberman, M.A. {\it Regular and Chaotic Dynamics} (Springer, Berlin. 1992). 
\bibitem{venegeroles_2009} R. Venegeroles, Phys. Rev. Lett. {\bf 101}, 054102 (2008); R. Venegeroles, Phys. Rev. Lett. {\bf 102}, 064101 (2009).
\bibitem{zaslavsky_2002} G.M. Zaslavsky, Phys. Rep., {\bf 371}, 461 (2002).
\bibitem{meiss_1985} J. D. Meiss and E. Ott, Phys. Rev. Lett. {\bf 55}, 2741 (1985).
\bibitem{zaslavsky_2002a} G.M. Zaslavsky, Physica D, {\bf 168-169}, 292 (2002).
\bibitem{altmann_2008} E. G. Altmann and T. Tel, Phys. Rev. Lett. {\bf 100}, 174101 (2008).
\bibitem{sanjuan_2008} J. M. Seoane, M. A.F. Sanju\'an, Phys. Letts. A {\bf 372}, 110 (2008).
\bibitem{altmann_2010}E. G. Altmann and A. Endler, Phys. Rev. Lett. {\bf 105}, 244102 (2010).
\bibitem{tamas_lai_book} Y.-C. Lai and T.  T\'el {\it Transient Chaos Complex Dynamics on Finite-Time Scales} (Springer, New York. 2010).
\bibitem{alus_2014} O. Alus, S. Fishman J. D. Meiss,  arXiv:1410.7648v1 [nlin.CD], 2014.
\bibitem{meiss_1993}R.W. Easton, J.D. Meiss and S. Carver, Chaos {\bf 3} 153 (1993). 
\bibitem{lopes_2012} S. R. Lopes, J. D. Szezech Jr., R. F. Pereira, A. A. Bertolazzo, and R. L. Viana, Phys. Rev. E, {\bf 86}, 016216 (2012).
\bibitem{balbiano_1994} A. Babiano, G. Boffetta, A. Provenzale, and A. Vulpiani, Phys. Fluids {\bf 6}, 2465 (1994).
\bibitem{tel_2005} T. T\'el, A.  de Moura, C. Grebogi and G. K\'arolyi, Phys. Rep. {\bf 413}, 91 (2005). 
\bibitem{szezech_2012}J. D. Szezech Jr., I. L. Caldas, S. R. Lopes, P. J. Morrison, and R. L. Viana, Physical Review E {\bf 86}, 036206(2012).
\bibitem{Efthymiopoulos_1999} C. Efthymiopoulos, G Contopoulos, N Voglis, Celestial Mechanics and Dynamical Astronomy, {\bf 73}, 221 (1999). M. Harsoula, C. Kalapotharakos, G. Contopoulos, MNRAS, {\bf 411}, 1111 (2011).
pp 221-230
\bibitem{bunimovich_2012} L. A. Bunimovich and L. V. Vela-Arevalo, Chaos {\bf 22}, 026103 (2012).
\bibitem{mackay_1984} R. S. Mackay, J. D. Meiss and I. C. Percival, Physica {\bf 13D}, 55 (1984). 
\bibitem{guckenheimer_2002} J. Guckenheimer and P. Holmes, {\it Nonlinear Oscillations, Dynamical Systems, and Bifurcations of Vector Fields} (Springer, New York, 2002).
\bibitem{grebogi_1990} C. Grebogi {\it et al.}, Phys. Rev. Lett., {\bf 65}, 1527 (1990).
\bibitem{ginelli_2007} F. Ginelli {\it et al.}, Phys. Rev. Lett., {\bf 99}, 130601 (2007).
\end{thebibliography}
\end{document}